\documentclass[prb,twocolumn,showpacs,amsmath,amssymb,floatfix]{revtex4}
\usepackage{graphicx}

\begin{document}

\title{Superconductivity induced by spark erosion in ZrZn$_2$}
\author{E. A. Yelland}
\author{S. M. Hayden}
\author{S. J. C.~Yates}
\affiliation{H.~H. Wills Physics Laboratory, University of
Bristol, Tyndall Avenue, Bristol BS8 1TL, United Kingdom}

\author{C. Pfleiderer}
\altaffiliation{Present address: Physik-Department E21, Technische
Universit\"{a}t M\"{u}nchen, D-85747 Garching, Germany}
\affiliation{Physikalishes Institut, Universitat Karlsruhe,
D-76128 Karlsruhe, Germany}
\author{M. Uhlarz}
\author{R. Vollmer}
\author{H. v.\ L\"{o}hneysen}
\affiliation{Physikalishes Institut, Universitat Karlsruhe,
D-76128 Karlsruhe, Germany}
\author{N. R. Bernhoeft}
\affiliation{DRFMC-SPSMS, CEA Grenoble, F-38054 Grenoble, Cedex 9, France}
\author{R. P. Smith}
\author{S. S. Saxena}
\affiliation{Cavendish Laboratory, University of Cambridge,
Madingley Road, Cambridge CB3 0HE, United Kingdom}
\author{N. Kimura}
\affiliation{Center for Low Temperature Science, Tohoku
University, Sendai, Miyagi 980-8578, Japan}

\date{\today}

\begin{abstract}
We show that the superconductivity observed recently in the weak itinerant
ferromagnet ZrZn$_2$ [C. Pfleiderer \textit{et al.}, Nature (London)
\textbf{412}, 58 (2001)] is due to remnants of a superconducting layer
induced by spark erosion. Results of resistivity,
susceptibility, specific heat and surface analysis measurements on
high-quality ZrZn$_2$ crystals show that cutting by spark erosion leaves a
superconducting surface layer. The resistive superconducting transition is
destroyed by chemically etching a layer of 5\,$\mu$m from the sample. No
signature of superconductivity is observed in $\rho(T)$ of etched samples at
the lowest current density measured, $J=675$\,Am$^{-2}$, and at $T\geq 45$\,mK.
EDX analysis shows that spark-eroded surfaces are strongly Zn depleted. The
simplest explanation of our results is that the superconductivity results from
an alloy with higher Zr content than ZrZn$_2$.

\end{abstract}

\pacs{75.50.Cc, 74.70.Ad, 74.25.Fy, 74.70.-b}


\maketitle

\section{\label{sec:Intro} Introduction}

Ferromagnetism in the cubic Laves compound ZrZn$_2$ was first
discovered by Matthias and Bozorth \cite{Matthias58} in 1958.
Since this time, ZrZn$_2$ has attracted considerable theoretical
and experimental attention. In particular, some authors have
suggested that metals which are close to a ferromagnetic
instability at low temperatures, such as ZrZn$_2$, may exhibit
magnetically mediated $p$-wave superconductivity
\cite{Leggett78,Fay80}. In principle, ZrZn$_2$ is an ideal
material in which to search for such $p$-wave superconductivity
because it can be driven into the paramagnetic state by
application of relatively modest pressures \cite{Smith71}.
However, until recently, experiments failed to find any evidence
for superconductivity in ZrZn$_2$
[\onlinecite{Falge66,Wu80,Cordes81,Huang81}].

Signatures of weak superconductivity were recently reported in the
magnetic and transport properties of ZrZn$_2$
[\onlinecite{Pfleiderer01a}]. In this paper, we show that spark
erosion, a standard procedure for cutting metallic samples, can
produce a superconducting surface layer on ZrZn$_2$ at ambient
pressure. The samples used in the present work are from the same
ingot as those used in Ref.~\onlinecite{Pfleiderer01a} and have
residual resistance ratios (RRR's) as high as 105.

\section{\label{sec:Exp} Experimental Details}

ZrZn$_2$ melts congruently at $1180^\circ$C
\cite{Elliot65,Massalski86}. At this temperature zinc has a vapor
pressure of about 10~bars and is an aggressive flux. Thus we chose
to grow ZrZn$_2$ by a directional cooling technique
\cite{Schreurs89}. Stoichiometric quantities of high-purity
zone-refined Zr (99.99\%, Materials Research MARZ grade) and Zinc
(99.9999\%, Metal Crystals) were loaded into a Y$_2$O$_3$
crucible. The total charge used was 4.2\,g. The crucible was
sealed inside a tantalum bomb which was closed by electron beam
welding under vacuum.  The assembly was heated to $1210^\circ$C
and then cooled through the melting point at
$2^\circ$C\,hr$^{-1}$.  The ingot was then annealed by cooling to
$500^\circ$C over a period of 72 hr. This method produced single
crystals of volumes up to approximately 0.4~cm$^{3}$. Single
crystals produced in this way had residual resistivities as low as
$\rho_0$= 0.53\,$\mu \Omega \mathrm{cm}$ corresponding to RRR~=
$\rho(293 K)/\rho(T \rightarrow 0)$\ =105. This corresponds to a
quasiparticle mean free path $\ell =1500$\,\AA\ (assuming a Fermi
surface area $S_F=1.9\times10^{21}$\ m$^{-2}$ , as given by
band-structure calculations \cite{Yates03}).

\begin{figure*}
\begin{center}
\includegraphics[width=0.95\linewidth,clip]{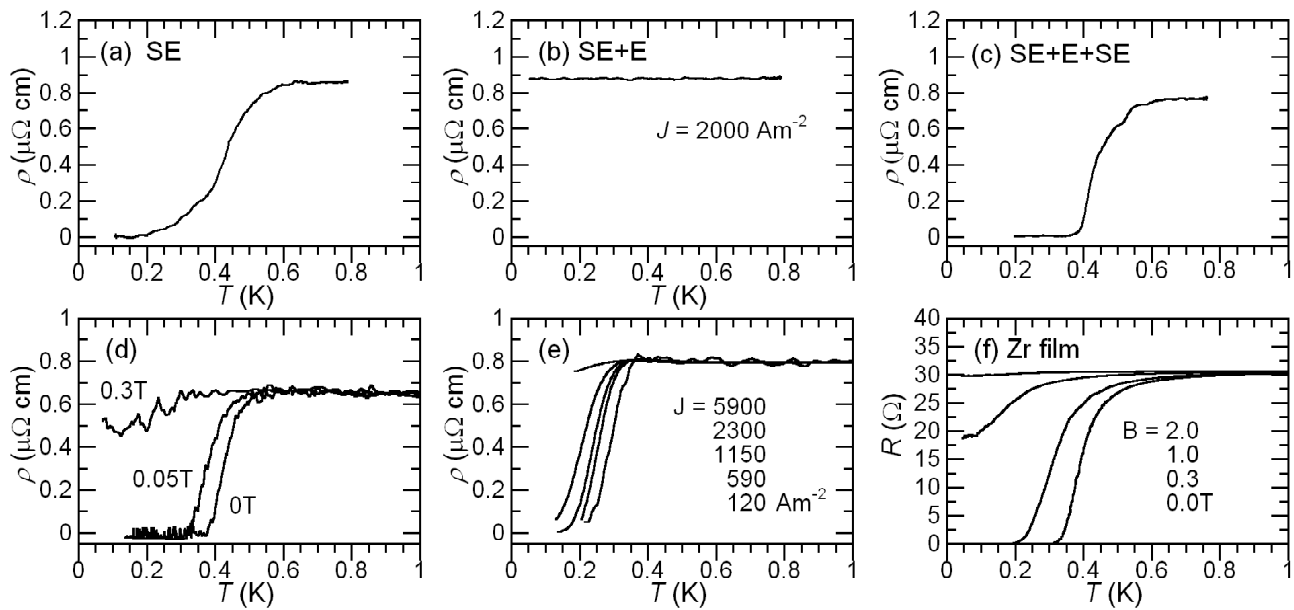}
\end{center}
\caption{ \label{f:rho_T} (a)-(e) The temperature dependence of
the resistivity of ZrZn$_2$ under various conditions and for
different sample treatments. The current density $J$ is calculated
using the bulk cross-sectional area of the sample. (a) As grown
sample cut by spark erosion ($B=0$\,T, $J=2000$\,Am$^{-2}$); (b)
as (a) followed by HF/HNO$_{3}$ etching ($B=0$\,T,
$J=2000$\,Am$^{-2}$); (c) as (b) followed by further spark erosion
($B=0$\,T, $J=2000$\,Am$^{-2}$); (d) The effect of applied field
on $\rho(T)$ of a spark-eroded sample. (e) The dependence on
current density of $\rho(T)$ of a spark-eroded sample. (f)
$\rho(T,B)$ for a zirconium film with a magnetic field applied in
plane of film.}
\end{figure*}

All experiments reported here were measured at ambient pressure.
Resistivity measurements were made using a standard a.c.\
technique using a Brookdeal 9433 low-noise transformer and SR850
digital lock-in amplifier. Most measurements were made at a
frequency $f$=2\,Hz. Sample contacts were made with Dupont 4929
Ag/epoxy. Heat capacity measurements were made using a long-pulse
technique in which the sample was mounted on a silicon platform
connected to a temperature-controlled stage by a thin copper wire.

Energy dispersive X-ray (EDX) analysis of sample surfaces was
performed on a Jeol JSM-5600LV scanning electron microscope using
a 20\,kV incident electron beam. In order to make quantitative
composition estimates, we recorded EDX spectra from sample
surfaces and from elemental standards under identical conditions.
All surfaces analyzed, except the spark-eroded surface, were
prepared by polishing with 0.1\,$\mu$m diamond lapping film in
order to minimize errors due to geometrical effects.
\section{Results}

\subsection{\label{sec:SC_res}Resistivity}
Fig.~\ref{f:rho_T} shows the temperature dependence of the
resistivity for ZrZn$_2$ under various conditions and for
different sample treatments. In order to make resistivity
measurements, bar-shaped samples were cut from the ingot by spark
erosion using Mo wire. Fig.~\ref{f:rho_T}(a) shows the $\rho(T)$
curve for a sample with all surfaces produced by spark erosion. A
superconducting transition is observed with an onset temperature
($T_{\mathrm{SC}}$) of about 0.6\,K. To test whether the
superconductivity is a bulk property or a property of the
spark-eroded surfaces, we then etched the sample used in
Fig.~\ref{f:rho_T}(a) in a solution containing 12 parts by volume
of 69\% HNO$_3$, 5 parts 48\% HF and 1000 parts H$_2$O for 1
minute. This removed 5\% of the sample mass, corresponding to a
surface layer 5\,$\mu$m thick. Fig.~\ref{f:rho_T}(b) shows the
resistivity measured after etching: the superconducting transition
has been removed.  The sample was then spark-cut along its length
to give two pieces, each having one spark-eroded surface. Panel
(c) shows $\rho(T)$ for one of these. The superconducting
transition has been restored. These results were obtained with the
voltage contacts on the spark-eroded surface, but identical
behavior was observed when the experiment was repeated on another
sample with the same treatment history and voltage contacts were
placed on the etched surface opposite to the spark-eroded one. On
this occasion, Cu wire was used as the spark-cutter electrode so
contamination from the wire is excluded as a cause of surface
superconductivity. The disappearance of superconductivity after
etching and its subsequent reappearance after spark-cutting was
reproduced in another sample. In well etched samples no sign of
superconductivity was observed at the lowest current density
measured $J=675$\,Am$^{-2}$ and the lowest temperature $T=45$\,mK.

We also investigated the magnetic field dependence of the
superconducting transition in spark-eroded ZrZn$_2$ samples.
Fig.~\ref{f:rho_T}(d) shows the resistive transitions measured
with a magnetic field applied perpendicular to the current and
parallel to one of the spark-eroded surfaces; the superconducting
anomaly persists to fields $\mu_0 H > 0.3$\,T. At first glance
this implies a surprisingly high critical field to critical
temperature ratio but as we discuss later this is easily explained
by the reduction of the Meissner screening in a layer that is thin
compared to the superconducting penetration depth.
Fig.~\ref{f:rho_T}(f) shows resistivity results for a Zr film
which will be compared with our results on ZrZn$_2$ below. The
film was produced by evaporation of Zr wire on to a glass
substrate under a vacuum of $\approx 10^{-5}$\,torr. The film
thickness is estimated as $\sim 500$\,nm. Superconductivity
persists with in-plane magnetic fields up to $\sim 1$\,T.

\subsection{\label{sec:SC_sus}Susceptibility}
\begin{figure}
\begin{center}
\includegraphics[width=0.95\linewidth,clip]{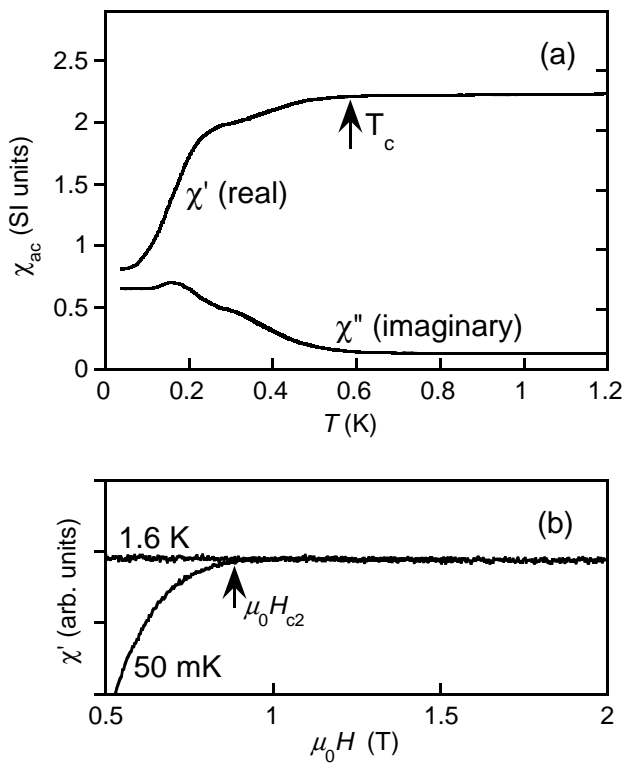}
\end{center}
\caption{ \label{f:zrzn2_chi_T_B} (a) The temperature dependence
of the a.c.\ susceptibility of ZrZn$_2$ (sample CY).  Measurements
were made with a modulation field of amplitude $b$\ $=1.2~\mu$T
and frequency $f$\ $=1.17$~kHz. The large susceptibility in the
normal state is due to the presence of ferromagnetism. (b) The
field dependence of the a.c.\ susceptibility measured with the
same modulation amplitude and frequency as the main figure.}
\end{figure}
Fig.~\ref{f:zrzn2_chi_T_B} shows a.c.\ susceptibility measurements
made on a spark-eroded sample of ZrZn$_2$.  The susceptometer was
calibrated by measuring the superconducting transition of an
indium sample at low frequencies ($f$=1.5--5\,Hz). The data have
not been corrected for the effects of the demagnetization field.
Fig.~\ref{f:zrzn2_chi_T_B}(a) shows that both the real and
imaginary components of the susceptibility ($\chi^{\prime}$ and
$\chi^{\prime \prime}$) are large in the temperature range $0.6<T<
1.2$~K, due to the alignment of ferromagnetic domains by the
applied field. Below the onset temperature of the spark
erosion-induced superconductivity $T_{\mathrm{SC}} \approx$\
0.6~K, the real part of the susceptibility begins to drop and the
imaginary part starts to increase.  There is a large drop in the
susceptibility $\Delta \chi^{\prime} \approx 2$, however full
Meissner screening [$\chi^{\prime} = -1/(1-N)$] is not observed.
We estimate the demagnetization factor of the sample to be $N
\approx 0.2 $. It appears that spark erosion induces a thin
superconducting layer which partially shields the ferromagnetic
core. Fig.~\ref{f:zrzn2_chi_T_B}(b) shows the a.c.\ susceptibility
measured in the presence of d.c.\ magnetic fields up to 2\,T for
$T$=0.05\,K and $T$=1.6\,K.  The difference between the 0.05\,K
and 1.6\,K curves shows that some parts of the sample have
critical fields up to 0.9\,T.

\subsection{\label{sec:SC heat} Heat Capacity}
\begin{figure}
\begin{center}
\includegraphics[width=0.85\linewidth,clip]{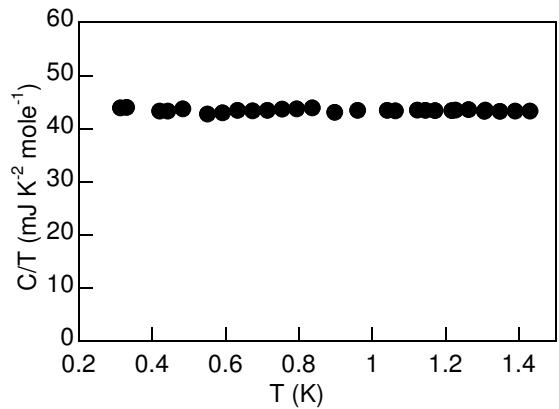}
\end{center}
\caption{ \label{f:zrzn2_C_T} The specific heat capacity of
ZrZn$_2$ (sample CS) in zero applied field measured by a
relaxation method. The absence of a heat capacity anomaly at
$T_\mathrm{SC}$ strongly suggests that the bulk is not
superconducting above 0.3\,K.}
\end{figure}
One of the most direct signatures of bulk superconductivity is the
specific heat anomaly. Fig.~\ref{f:zrzn2_C_T} shows the specific
heat capacity, plotted as $C(T)/T$, for a sample (CS) cut from a
region of the ingot next to that used for the resistivity and
susceptibility measurements. As in previous work
\cite{Pfleiderer01a}, no specific heat capacity anomaly is
observed, strongly suggesting the absence of bulk
superconductivity above $T=0.3$\,K in this sample.

\subsection{EDX Analysis}

\begin{figure}
\begin{center}
\includegraphics[width=0.95\linewidth,clip]{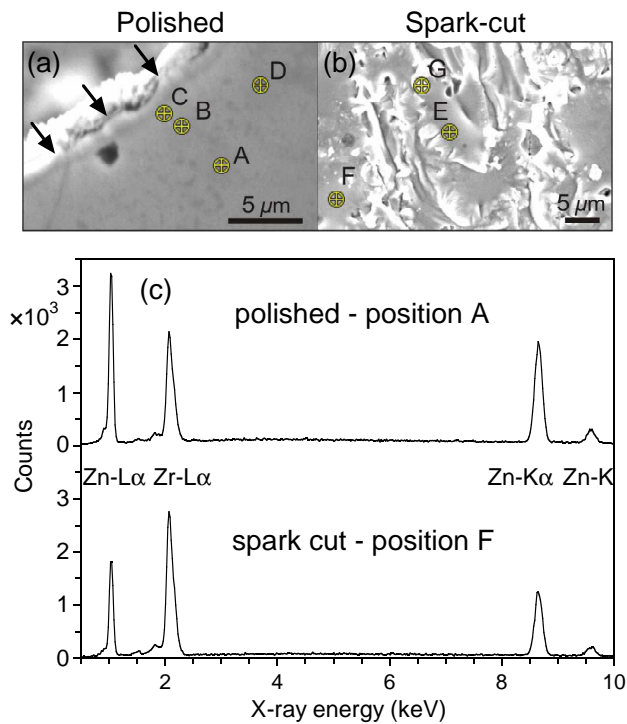}
\end{center}
\caption{Panel (a): SEM image of a polished ZrZn$_2$ sample. The
polished face is normal to the incident electron beam; the sample
edge, indicated by arrows, is defined by a spark-cut surface that
is approximately parallel to the beam. Thus the EDX analysis at
position C probes a region $1-2\,\mu$m below the spark-cut face.
Panel (b): SEM image of a spark-cut sample. Panel (c): examples of
raw EDX spectra obtained on spark-cut and polished samples,
showing Zn depletion of the spark-cut surface.} \label{f:SEMEDX}
\end{figure}

We have demonstrated above that spark erosion of ZrZn$_2$ induces
a superconducting layer. In order to determine the nature of the
changes in the surface layer that cause the behavior shown in
Figs.~\ref{f:rho_T}(a-c), we performed spot-mode EDX analysis on a
spark-cut surface of a superconducting ZrZn$_2$ sample
[Fig.~\ref{f:SEMEDX}(b)] and on a polished surface
[Fig.~\ref{f:SEMEDX}(a)]. The latter was exposed by cleaving a
sample such that the cleave plane was perpendicular to a spark-cut
surface; the cleaved surface was then polished to ensure that it
was flat and perpendicular to the incident electron beam. We
estimate that the spatial resolution of the EDX probe is $\sim
1\,\mu$m in all directions \cite{ScottBook}.

\begin{table}
\begin{ruledtabular}
\begin{tabular}{l|ll|ll}
peaks analysed& \multicolumn{2}{c|}{Zr-L$\alpha$/Zn-K$\alpha$}&\multicolumn{2}{c}{Zr-L$\alpha$/Zn-L$\alpha$}\\
\colrule
& Zr(\%)&Zn(\%)&Zr(\%)&Zn(\%) \\
\colrule
Position&\multicolumn{4}{c}{polished}\\
\colrule A&31.5&68.5&33.0&67.0 \\
B&31.6&68.4&33.0&67.0 \\
C&32.2&67.8&33.3&66.7 \\
D&31.7&68.3&32.8&67.2 \\
\colrule
&\multicolumn{4}{c}{spark-cut}\\
\colrule E&49.1&50.9&42.2&57.8\\
F&47.3&52.7&48.8&51.2\\
G&71.1&28.9&67.7&32.3\\
\end{tabular}
\end{ruledtabular}
\caption{\label{table:EDXresults} Atomic concentration results of
spot EDX analysis on polished and spark-cut samples of ZrZn$_2$.
The position labels correspond to those in Fig.~\ref{f:SEMEDX}. A
ZAF correction procedure using pure element standards was used to
calculate atomic concentrations. Two characteristic Zn peaks,
Zn-K$\alpha$ (8.6\,keV) and Zn-L$\alpha$ (1.0\,keV), were used to
give two distinct concentration estimates; the agreement of the
two estimates shows that the roughness of the spark-cut surface
does not prevent a quantitative analysis. We conclude that the
composition of the spark-cut surface varies but that it is always
Zn depleted.}
\end{table}

The EDX spectra taken on nominally ZrZn$_2$ surfaces [see
Fig.~\ref{f:SEMEDX}(c)] contain three characteristic X-ray
emission peaks that provide useful information about sample
composition: Zn-L$\alpha$ (1.0\,keV), Zr-L$\alpha$ (2.0\,keV) and
Zn-K$\alpha$ (8.6\,keV). The spectra in Fig.~\ref{f:SEMEDX} show
at a glance that the Zr peak is enhanced relative to both Zn peaks
on the spark-cut surface compared to the polished surface,
suggesting that spark erosion causes Zn depletion at the surface.
We verified this by quantitative EDX analysis using spectra
obtained both on ZrZn$_2$ samples and on pure element standards in
identical conditions; for each peak $i$, the ratio $k_i$ of peak
area in the ZrZn$_2$ spectrum to that in the spectrum of the pure
element standard was found. To a first approximation, the mass
concentration of each element in the test sample is given by these
standard-normalized peak areas $k_i$. However, it is well known
that corrections \cite{ScottBook} must be applied to take account
of the way in which the composition of the test sample affects
e.g.\ the electron beam penetration and the absorption of
generated X-rays. Therefore we used a standard iterative ZAF
(atomic number, absorption and fluorescence) correction procedure
as implemented in the CITZAF package \cite{Armstrong95} to give
accurate atomic concentrations, which are summarized in
Table~\ref{table:EDXresults}. The quantitative result for the
polished surface is in good agreement with the nominal atomic
concentration (i.e.\ 33.3\% Zr, 66.7\% Zn) and the small
difference is within the error arising from imperfect ZAF
correction as discussed later. There is no dependence of the
composition on distance from the spark-cut surface in the results
shown for the polished surface, implying that any compositional
change extends to a depth $\leq1-2\,\mu$m. The results for the
spark-cut surface show that the composition varies as a function
of position but that it is always Zn-depleted, and that it
includes regions that are close to Zr$_2$Zn in composition. Atomic
concentrations shown in Table~\ref{table:EDXresults} are
normalized to 100\% but we note that no significant contaminant
peaks are present in any of the spectra used to produce these
results.  For each sample spectrum two different composition
estimates were obtained, the first using only the Zr-L$\alpha$ and
Zn-K$\alpha$ peaks and the second using Zr-L$\alpha$ and
Zn-L$\alpha$. The presence of two characteristic Zn X-ray peaks of
very different energy provides a check on our correction procedure
because the correction factor is much larger for Zn-L$\alpha$ than
for Zn-K$\alpha$. The excellent agreement between the two
estimates for the polished surface shows that the correction
procedure provides accurate results in this ideal geometry. The
small discrepancy (at the $1\%$ atomic concentration level) may
arise, for example, from the correction model used or from a
slight tilt of the sample surface relative to the electron beam.
Importantly, the results for the spark-cut surface obtained using
the different Zn peaks are also in good agreement. The discrepancy
is larger than in the case of the polished surface due to the
effects of surface roughness on the X-ray intensity corrections.
However, it is still small compared to the dramatic Zn depletion
observed. Thus, we have shown that cutting ZrZn$_2$ by spark
erosion causes the formation of a Zn-depleted surface layer of
thickness $\leq 1-2\,\mu$m.

\section{\label{sec:discussion} Discussion}

We have demonstrated that ZrZn$_2$ is very susceptible to surface
damage caused by spark erosion. The spark erosion process causes
the surface layer to be depleted of Zn. Removal of the remaining
Zr-rich surface layer requires an HF-based etch or
electropolishing. In the course of this work we have found that
un-etched spark-cut regions can easily be left on the surface if
they are protected, for example, by organic material such as
remnants of Ag/epoxy paint contacts. The resulting samples show
resistive downturns like those observed in
Ref.~\onlinecite{Pfleiderer01a}. Although the resistivity
measurements reported in Ref.~\onlinecite{Pfleiderer01a} were made
with contacts on cleaved surfaces, the current path included
remnants of spark-cut surfaces.

Our EDX measurements show that although the composition of the
spark-cut surface varies in space, it is always more Zr-rich than
ZrZn$_2$. The high mobility of Zn in the surface layer of ZrZn$_2$
is likely to be connected with the low melting point of Zn, $T_m$
= 419.6$^\circ$C. No regions of elemental Zr or Zn were observed
at the 1\,$\mu$m resolution of the EDX probe.  Thus the simplest
explanation for our results is that the observed downturns in
$\rho(T)$ and $\chi(T)$ are due to a surface layer of a
superconducting alloy, with higher Zr content than ZrZn$_2$, that
is created by spark erosion.  It is unlikely that spark cutting
produces pure Zr because of the high solubility of Zn in Zr
[\onlinecite{Elliot65,Massalski86}]. Other scenarios are also
possible. One intriguing possibility is that spark erosion creates
strained layers near the surface which are superconducting
\cite{Kimura04}.

A feature of the spark-erosion-induced superconducting layer in
ZrZn$_2$\ is its large critical field to critical temperature
ratio.  It is well known that the critical field of a
superconducting sample is enhanced with respect to the bulk
thermodynamic critical field $B_c$\ when the sample is
sufficiently small in at least one direction $\perp \mathbf{B}$\
to allow penetration of magnetic flux \cite{Tinkham75}. For
example, Al films of thickness $250$\,\AA\ can show
$T_\mathrm{SC}\approx1.7$\,K and a critical field of 1.9\,T
[\onlinecite{Strongin66}]. Fig.~\ref{f:rho_T}(f) demonstrates this
effect in a film of Zr: the critical field of the film is $\sim
1$\,T (close to the paramagnetic limit $B_c^\mathrm{para}=1.84\,
T_{c}$\ [\onlinecite{Clogston62,Chandrasekhar62}]) compared to
that of bulk Zr, $B_{c} \approx$\ 0.0047~T
[\onlinecite{CRCHandbook}]. A similar enhancement is likely to
occur in spark-eroded ZrZn$_2$ at any regions of the
superconducting layer that are thin compared to the penetration
depth.

We now comment on the pressure dependence of $T_\mathrm{SC}$ and
$T_\mathrm{FM}$. Ref.~\onlinecite{Pfleiderer01a} suggested that
superconductivity and ferromagnetism vanished simultaneously at a
critical pressure $p_c\sim 20$\,kbar. More detailed measurements
of $T_\mathrm{FM}(p)$ [\onlinecite{Uhlarz04}] have now shown that
ferromagnetism in fact disappears in a first order transition at a
lower pressure $p_c(\mathrm{FM})=16.5$\,kbar. Unfortunately no new
data are available for $T_\mathrm{SC}(p)$, but the results in
Ref.~\onlinecite{Pfleiderer01a} show that
$13<p_c(\mathrm{SC})<22$\,kbar. Thus the two $p_c$'s lie close to
each other, but further measurements of $T_\mathrm{SC}(p)$ would
be needed to establish whether the superconductivity of the
surface layer is related to ferromagnetism in the bulk.

\section{Conclusions}
In summary, spark erosion induces a superconducting layer in
ZrZn$_2$. If this surface layer is removed by chemical etching the
resistive superconducting transition disappears. EDX analysis of
spark-cut surfaces shows that they are Zn depleted. The simplest
explanation for the induced superconductivity is that it is due to
a change in chemical composition caused be the spark erosion. It
remains to be seen whether higher quality ZrZn$_2$ is
superconducting at ambient pressure and above.

\section{Acknowledgements}
We are grateful to A. Carrington and G. G. Lonzarich for useful
discussions.

\bibliography{zrzn2_spark_PRB_condmatfinal}


\end{document}